IAC-16-D1.2.1 (34366)

# Automated Design of CubeSats and Small Spacecrafts

Himangshu Kalita[a], Jekanthan Thangavelautham[b*]

[a] *School of Energy, Matter and Transport Engineering, Arizona State University, Tempe, Arizona 85281, United States of America*
[b] *School of Earth and Space Exploration, Arizona State University, Tempe, Arizona 85281, United States of America*,
jekan@asu.edu
* Corresponding Author

**Abstract**

The miniaturization of electronics, sensors and actuators has enabled the growing use of CubeSats and sub-20 kg spacecraft. Their reduced mass and volume has the potential to translate into significant reductions in required propellant and launch mass for interplanetary missions, earth observation and for astrophysics applications. There is an important need to optimize the design of these spacecraft to better ascertain their maximal capabilities by finding optimized solution, where mass, volume and power is a premium. Current spacecraft design methods require a team of experts, who use their engineering experience and judgement to develop a spacecraft design. Such an approach can miss innovative designs not thought of by a human design team. In this work we present a compelling alternative approach that extends the capabilities of a spacecraft engineering design team to search for and identify near-optimal solutions using machine learning. The approach enables automated design of a spacecraft that requires specifying quantitative goals, requiring reaching a target location or operating at a predetermined orbit for a required time. Next a virtual warehouse of components is specified that be selected to produce a candidate design. Candidate designs are produced using an artificial Darwinian approach, where fittest design survives and 'reproduce', while unfit individuals are culled off. Our past work in space robotic has produced systems designs and controllers that are human competitive. Finding a near-optimal solution presents vast improvements over a solution obtained through engineering judgment and point design alone. Through this design approach, we evaluate a LEO- deployed 6U CubeSat that needs to generate a required average power. The approach identifies credible solution that will need further study to determine its implementation feasibility. The approach shows a credible pathway to identify and evaluate many more candidate designs than it would be otherwise possible with a human design team alone.

## 1. Introduction

Space systems perform important tasks, including planetary exploration, astronomical observations, earth observation, technology demonstrations and communication in space. A typical spacecraft may undergo extreme changes in temperature, cosmic and solar particle induced radiation, withstand the vacuum of space and handle launch shock and vibrations. Designing a spacecraft is a long, expensive endeavour, where a system is tailor designed for a specific mission at hand. A new design approach is required that shortens the spacecraft design process.

Rapid technology advancement in the CubeSat and small satellite industry has led to standards defining mass, volume and launch specifications. There has been rapid development of modular, interchangeable spacecraft components, including computer electronics, science instruments, power supplies, communication device to name a few. In this paper, we propose automated design of a spacecraft utilizing modular, interchangeable components. In our approach, a numerical goal function is specified, along with constraints. An Evolutionary Algorithm (EA) is used to generate a population of candidate solutions, where the fittest individual mate and mutate, while unfit individuals are culled (Fig. 1). The candidate population is evolved for hundreds of generations until the fittest individual in the population meets a desired performance metric.

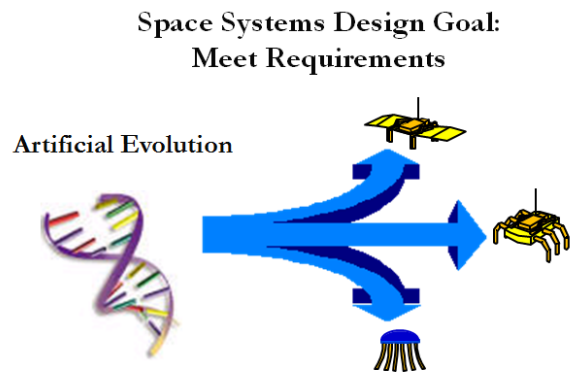

Fig. 1. Automated Design of Space Systems.

With this approach, finding sets of optimal designs that represent the trade-offs among conflicting objectives, such as mass, cost, and performance, can be very useful in making informed system design decisions. Automated design of engineering systems is not new. Computationally derived evolutionary designs have shown competitive advantages over human created designs in terms of performance, creativity and robustness. Researchers have been investigating evolutionary design and optimization for years and several satellite subsystems have been investigated including antennas [1], power system [2] and low-thrust orbit transfer [3]. This process also has been successfully applied to design of mobile robots [6] and water desalination systems [7].

Automated design is in sharp contrast to current spacecraft design methods that require a team of experts, who use their engineering experience and judgement to develop a spacecraft design. The initial identification of candidate designs is based on individual judgement and is often limited to dozens of designs. It is time and labor intensive and require significant expertise and experience. Wrong assumptions may lead to a sub-optimal design or worse an intractable solution.

There is typically no systematic approach to evaluate the whole design space that can meet the defined goals and satisfy the constraints. The principle limiting factor is the ability for a team to fully evaluate a candidate spacecraft design and quantitatively determine its strengths and limitations. Such an approach can miss innovative designs not thought of by the design team. Evolutionary design techniques can overcome these limitations by searching the design space and automatically finding effective solutions that would ordinarily not be found. In the following sections, we present background and related work on automated design (Section 2), description of the automated design (Section 3), results and discussion (Section 4), followed by conclusions and future work (Section 5).

**2. Background and Related Work**

Evolutionary Algorithms (EAs) are a stochastic search method that mimic the metaphor of natural biological evolution. It provides an approach to learning that is based loosely on Darwinian evolution. Evolutionary Algorithms operate on a population of potential solutions applying the principle of survival of the fittest to produce a solution. At each generation, a new population is created by the process of selecting individuals by their highest fitness in the problem domain and breeding them together using operators, namely crossover and mutation borrowed from natural genetics. In theory, this process leads to the evolution of populations of individuals that are better suited to their environment than the individuals that they were created from, just as in natural adaptation [4].

Evolutionary algorithms model natural processes, such as selection, crossover and mutation. Fig. 2 shows the structure of a simple Evolutionary Algorithm (EA). Evolutionary Algorithms have been used to automate the process of design in various fields spanning robotics, communication electronics and spacecraft power systems [10-12]. This approach produces near optimal solutions when successful, which presents a big improvement over a solution obtained through engineering judgement and point design. The solution obtained needs further study to determine its implementation feasibility.

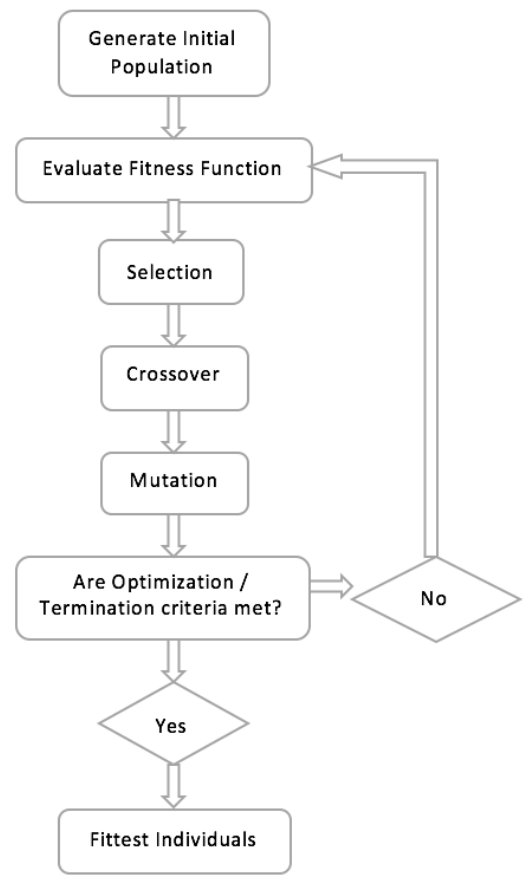

Fig. 2. Evolutionary Algorithm Structure

Our approach to the spacecraft design problem is modelled after the knapsack problem. Optimization of the knapsack problem is considered an NP hard problem. In the knapsack problem, the goal is fill a knapsack with as many books, so that the free volume is minimized and mass maximized. In our spacecraft design problem, the focus is effective packaging of components within a specified mass and volume constraint. Using this approach as we will show later, the gene specifies what components that are to be packaged inside the spacecraft. This keeps the gene and design process simple and produces results fast using desktop computers. There are also other approaches to automated design and this includes use of variable length generative coding schemes [8] that generates construction program to design the gene. Other bio-inspired approaches model morphogenesis [9].

| Subsystem Specifier | Type of Components | Type of SP | No. of Solar Panels | | | | | | | | | | | | | | |
|---|---|---|---|---|---|---|---|---|---|---|---|---|---|---|---|---|---|
| X | Y | T | B1 | B2 | B3 | B4 | E1 | E2 | E3 | E4 | E5 | E6 | E7 | E8 | V1 | V2 | V3 | V4 |
| Alphabets [C,O,X,B,A,P,T,S] | Alphabets [a,b,c,....,j] | Alphabets [b,e,v] | Integers [0 1] | | | | Integers [0 1 2 3] | | | | | | | | Integers [0 1 2 3] | | | |

Fig. 3. Description of Gene of CubeSat

## 3. Automated Space Systems Design

In this paper we propose to use Evolutionary Algorithms to automate the process of design of CubeSat. We start with the design of a 6U in LEO. The population in EAs are often represented by bit strings, so that they can be easily manipulated by genetic operators such as crossover and mutation. In this paper the population is represented by strings of letters and numbers with each letter representing a component of the CubeSat and each number representing its location, orientation or count (see Fig. 3). The subsystem specifier X distinguishes the different components of the CubeSat namely Antenna, OBC, Structure, Battery, Reaction Wheels, Payload, Thruster and Solar Panels. Y determines the type of components in the satellite with different mass, cost, dimensions, performance parameter and manufacturer. T determines the type of solar panel used in the design. It differentiates between body mounted panels and deployable (edge mounted and vertex mounted) panels. The gene then describes the number of panels mounted on its body, edge and vertices.

The initial set of population is created randomly which then passes through a filter that filters out the CubeSat population with missing subsystem. The rest then passes through the fitness function. The fitness function defines the criterion for ranking potential populations and for probabilistically selecting them for inclusion in the next generation of population. It calculates the power produced by the respective designs, its mass and cost and then normalises it to a range of 0 to 1. For power 1 represents the highest power that the CubeSat is capable of producing and 0 represents the minimum power that it can produce. Similarly, for mass and cost 1 represents the minimum mass, cost and 0 represents the maximum mass, cost. The normalized fitness for power, mass and cost are then multiplied to calculate the overall fitness of the particular design. The designs are then ranked in descending order according to their overall fitness value. Based on the ranking top 50% individuals are selected from the entire population. The selected population then undergoes crossover. The crossover operator produces two new offspring from two parent strings, by copying selected part of the string from each parent. After crossover operator the population undergoes the mutation operator. The mutation operator produces small random changes to the string by choosing a single element of the string at random, then changing its value. The mutation operator produces offspring from a single parent. We have considered a mutation rate of 20% for the evolutionary algorithm.

After mutation, the fitness of the resultant population is checked and if the optimization/termination criteria are met, the algorithm stops and the fittest individuals are produced. If not the whole algorithm runs again until the fittest individuals are produced. In our case for optimizing the solar panel configuration, we have considered the different types of solar panels commercially available. Based on the power, cost, volume and mass requirement the EA provides the optimized configuration of the solar panels. Moreover, it calculates the capacity of battery to be used based on the power produced by the solar panels.

*3.1 Discipline Models*

This subsection describes the models for all the discipline in a 6U CubeSat for calculating the power produced in a LEO orbit.

*3.1.1 Orbit Dynamics*

The orbit-dynamics discipline computes the Earth-to-satellite and Earth-to-Sun position vector in ECI frame according to equation (1). The $J_2$ and $J_3$ terms were considered because of their effect to rotate orbit plane on a scale of months. The orbit equation was solved using a Runge-Kutta Method (ode45 in Matlab) [5].

$$\vec{\ddot{r}} = -\frac{\mu}{r^3}\vec{r} - \frac{3\mu J_2 R_e^2}{2r^5}\left[\left(1-\frac{5r_z^2}{r^2}\right)\vec{r} + 2r_z\hat{z}\right] - \frac{5\mu J_3 R_e^3}{2r^7}\left[\left(3r_z - \frac{7r_z^3}{r^2}\right)\vec{r} + \left(3r_z - \frac{3r_z^2}{5r_z}\right)r_z\hat{z}\right] \quad (1)$$

*3.1.2 Attitude Dynamics*

The attitude of the satellite is needed to be calculated to know which side of the satellite is facing the sun. At any given time instance, the attitude is determined by applying the rotations from the ECI frame to the actual body-fixed frame. We modelled only the reaction wheel for actuation. The required inputs are computed from the satellite's angular-velocity profile. We do this by applying conservation of angular momentum to the satellite and reaction wheel system, expressed by setting the time derivative of the total angular momentum to zero according to equation (2) [5].

$$\dot{\vec{L}} = J_B \cdot \dot{\vec{\omega}}_B + \vec{\omega}_B \times (J_B \cdot \vec{\omega}_B) + J_{RW} \cdot \dot{\vec{\omega}}_{RW} + \vec{\omega}_B \times (J_{RW} \cdot \vec{\omega}_{RW}) = 0 \quad (2)$$

*3.1.3 Cell Illumination*

The cell-illumination discipline models the area of each solar panel that is exposed to the Sun, projected onto the plane normal to the Sun's incidence. First we calculate the line-of-sight variable LOS, which is essentially a multiplier for the exposed areas. It is 0 if the satellite is behind the Earth and 1 otherwise [5]. The exposed area is calculated by taking the dot product of the rotation matrix from ECI frame to the body-fixed frame with the Earth-Sun unit vector and multiplying with the total area along x, y and z axis and the LOS to to get the resultant exposed area.

*3.1.4 Solar Power*

The solar power produced at each time instant can be calculated by multiplying the solar constant with solar cell efficiency and total exposed area as shown in equation (3).

$$P = q_{sol} \times \eta_{sol} \times A_{exp} \quad (3)$$

**4. Results and Discussion**

The initial population is produced randomly and the fitness function for required power output, mass and cost defined the EA evolves the initial design to an optimized design that increases the power and reduces the mass and cost of the CubeSat. Fig. 4 shows the snapshots of the EA over 100 generations to produce the evolved design that can produce an output of 88W. Fig. 5 shows the average power produced in a LEO orbit by the fittest individual of each generation. It is clear that the EA evolves from the initial design of 33W to a final design of 81W. It can also be seen that the best design over 100 generations is able to produce 88W.

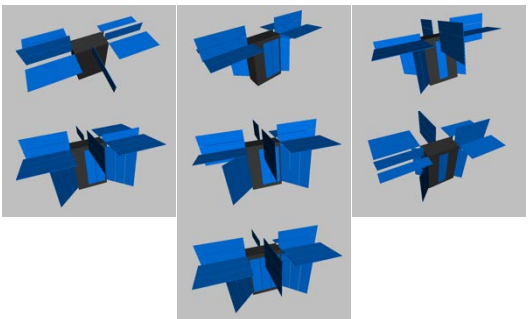

Fig. 4. Evolution of the design from 33W to 88W

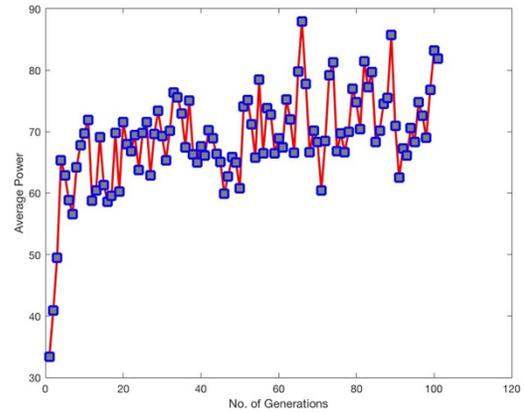

Fig. 5. Plot between Average Power and No. of Generations

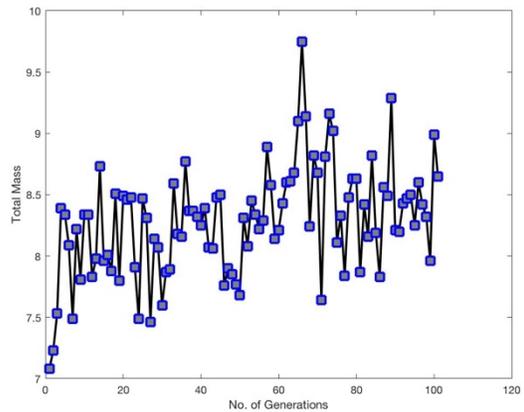

Fig. 6. Plot between Total Mass and No. of Generations

Fig. 6 and 7 shows the total mass and cost of the CubeSat designs through their evolution over 100 generations. The maximum mass and cost for a design that can produce a maximum of 90W is around 10.2 kg and $247,500 respectively. However, the EA optimises the three parameters and produces the fittest individuals. It is clearly evident from the three graphs (see Fig.5, Fig. 6 and Fig. 7) that the algorithm evolves the design such that the power output is maximised keeping mass and cost of the spacecraft minimised.

Fig. 8, 9 and 10 shows how the mean of average power, total mass and total cost of each generation changes. It can be clearly seen that the mean value for average power generation increases while that of total mass and cost decreases as the EA optimises the fitness function. The evolutionary process increases the mean fitness of each generation eliminating the unfit individuals and keeping the fittest individuals thus converging towards the population with the fittest individuals.

Preprint - 67th International Astronautical Congress (IAC), Guadalajara, Mexico, 26-30 September 2016.

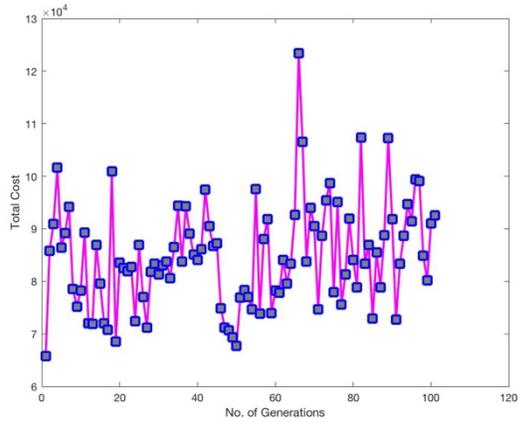

Fig. 7. Plot between Total Cost and No. of Generations

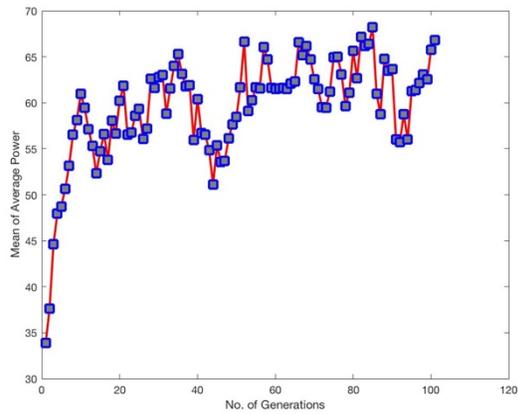

Fig. 8. Plot between Mean of Average Power and No. of Generations

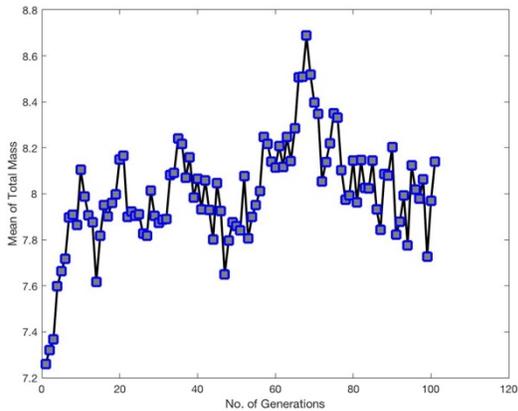

Fig. 9. Plot between Mean of Total Mass and No. of Generations

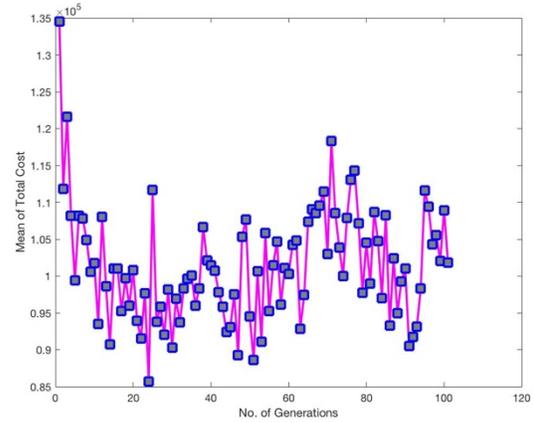

Fig. 10. Plot between Mean of Total Cost and No. of Generations

Fig. 11 shows the overall fitness (considering power, mass and cost together) of the best individuals of each generation. Here overall fitness is calculated within a range of 0-1, 1 representing the fittest individual and 0 representing the most unfit individual. It is clearly evident that the evolution starts from a design with minimum fitness and gradually evolves to a design with better fitness value.

This method of automatically designing CubeSats using Evolutionary Algorithm generates various near optimal designs which presents vast improvements over a solution obtained through engineering judgement and point design. These optimised designs will need further study to determine its implementation feasibility. This approach can also be used to design other subsystems of a CubeSat.

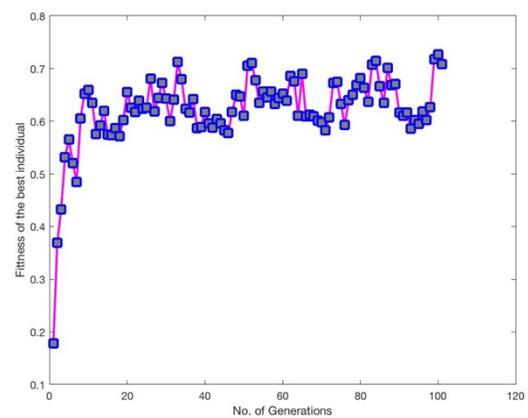

Fig. 11. Plot between Fitness of best Individual and No. of Generations

## 5. Conclusions

Evolutionary Algorithms have been successfully applied in finding near-optimal CubeSat designs. The approach has the potential for producing designs that maximize power while minimizing cost and mass. Our early experiments show promising results, with increasingly sophisticated designs that maximize power while reducing mass and cost. Usually trade studies are conducted by engineers to create mission concepts with different trade-off solutions for mass, cost, volume, performance and risk. Automated design using Evolutionary Algorithm speeds up the design process and provides with a better basis to make more detailed system architecture and design decisions with confidence.

**Appendix A (Data for Discipline Models)**

$(\vec{\cdot})$ = Vector
$(\hat{\cdot})$ = Unit vector
$A_{exp}$ = Exposed Area to Sun
$J$ = Mass moment of inertia matrix, kg.m$^2$
$L$ = Angular moment vector, kg.m$^2$/s
LOS = Satellite-to-Sun line of sight
P = Power, W
$r$ = Position vector norm, km
$\vec{r}$ = Position vector, km
$\omega$ = Angular velocity vector, 1/s
$J_2 = 1.08263 \times 10^{-3}$
$J_3 = -2.51 \times 10^{-6}$
$\mu = 398600.44$ km$^3$s$^{-2}$
$R_e = 6378.137$ km
$q_{sol} = 1.36 \times 10^3$ W/m$^2$
$\eta_{sol} = 0.284$